\documentclass[apl,aip,superscriptaddress,reprint]{revtex4-1}
\usepackage[dvips]{graphicx}
\usepackage{amssymb}
\usepackage{natbib}
\newcommand{\ped}[1]{\ensuremath{_{\rm #1}}}
\newcommand{\apex}[1]{\ensuremath{^{\rm #1}}}
\usepackage{color}
\usepackage{booktabs}

\begin{document}
\title{Strong dopant dependence of electric transport in ion-gated MoS\ped{2}}

\author{Erik Piatti}
\email{erik.piatti@polito.it}
\affiliation{Department of Applied Science and Technology, Politecnico di Torino, corso Duca degli Abruzzi 24, 10129 TO, Torino, Italy}

\author{Qihong Chen}
\author{Jianting Ye}
\email{j.ye@rug.nl}

\affiliation{Device Physics of Complex Materials, Zernike Institute for Advanced Materials, Nijenborgh 4, 9747 AG, Groningen, The Netherlands}

\begin{abstract}
We report modifications of the temperature-dependent transport properties of MoS\ped{2} thin flakes via field-driven ion intercalation in an electric double layer transistor. We find that intercalation with Li\apex{+} ions induces the onset of an inhomogeneous superconducting state. Intercalation with K\apex{+} leads instead to a disorder-induced incipient metal-to-insulator transition. These findings suggest that similar ionic species can provide access to different electronic phases in the same material.
\end{abstract}

\maketitle

Transition metal dichalcogenides are a fascinating class of layered materials, where different orders - such as superconductivity and charge-density waves - compete with each other and give rise to 
complex phase diagrams reminiscent of those of cuprates and iron pnictides \cite{KlemmBook2012,KlemmReview2015}. Intercalation by means of a wide range of compounds, both organic and inorganic, is a particularly powerful tool to tune the properties of these materials \cite{KlemmReview2015,LerfInorgChem1977,OnukiSynthMet1983}, 
resulting in superconducting compounds characterized by sharp transition temperatures and well-defined upper critical fields.


In recent years, ionic gating has been utilized to control the transport 
properties of a wide range of materials, including oxides 
\cite{UenoNatureMater2008,UenoNatureNano2011,BollingerNature2011,LengPRL2011,LengPRL2012,JeongScience13,JinSciRep2016}, 
metal chalcogenides \cite{YeScience2012,JoNanoLett2015,CostanzoNatureNano2015,YuNatureNano2015,ShiSciRep2015,XiPRL2016,
ShiogaiNaturePhys2015,LeiPRL2016,LeiPRB2017,LiNature2016,OvchinnikovNComms2016}, graphene \cite{EfetovPRL2010,YePNAS2011,GonnelliSciRep2015,PiattiAppSS2017} 
and other $2$-dimensional materials \cite{YeNatMater2010,SaitoScience2015,SaitoACSNano2015}, and even metals \cite{DagheroPRL2012,nakayama12,TortelloAppSS2013,ChoiAPL2014,
PiattiJSNV2016,PiattiPRB2017}.
Most of these result have been obtained within the electrostatic limit, \textit{i.e.} by 
only accumulating ions at the material surface and exploiting the ultrahigh 
electric field that develops in the electric double layer (EDL) \cite{UenoReview2014}. 
However, ionic gating of layered materials allows for a further degree of freedom 
in the technique, by exploiting the electric field to intercalate the ions between 
the van der Waals-bonded layers, thus allowing control over the properties 
of the entire bulk. This technique has already showcased its possibilities by allowing a robust 
control of the electronic ground state in TaS\ped{2} \cite{YuNatureNano2015}, 
MoTe\ped{2} \cite{ShiSciRep2015}, WSe\ped{2} \cite{ShiSciRep2015} and FeSe\cite{LeiPRB2017}. 
These studies mainly focused on the modulation of the bulk carrier density 
achieved via ion intercalation, without analyzing in detail the effects 
of different ionic species on the same ion-gated material. In principle, 
however, the choice of dopant ion may severely affect the properties 
of the intercalated phase, leading to ion-specific device behavior and 
possibly entirely different phase diagrams for the field-induced 
intercalated state. 

\begin{figure}
\includegraphics[width=0.8\columnwidth]{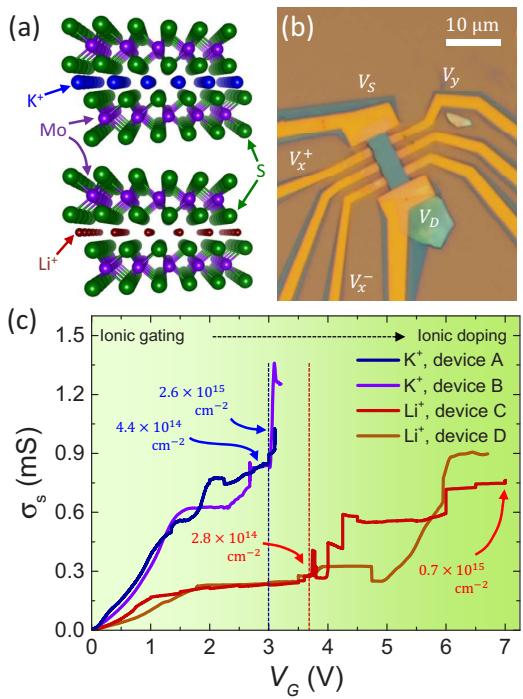}
\caption{(a) Ball-and-stick model of the MoS\ped{2} lattice with intercalated K\apex{+} (top panel) and Li\apex{+} (bottom panel) ions \cite{vesta2011}. (b) Optical micrograph of a MoS\ped{2} field-effect device before drop-casting the electrolyte. (c) Gate dependence of the sheet conductivity $\sigma_s$ at $T = 300$ K for both K\apex{+} (devices A and B) and Li\apex{+}-based electrolytes (devices C and D). Dashed lines indicate the corresponding threshold voltages for the onset of ion intercalation. The four values of densities correspond to the Hall carrier densities $n_H$ for devices A and C.}
\label{figure:1}
\end{figure}

Here, we tackle this issue by performing ionic gating experiments on archetypal 
layered semiconductor MoS\ped{2} using K\apex{+} and Li\apex{+} as dopant ions. 
MoS\ped{2} is known to undergo a series of insulator-to-metal-to-superconductor 
phase transitions both upon surface electrostatic carrier accumulation \cite{YeScience2012} 
and chemical intercalation with different ionic species \cite{WoollamMSE1977,SomoanoJCP1973}. 
We find that, for field-driven intercalation, this is the case only for the smaller Li\apex{+} ion 
(see the lower panel of Fig. \ref{figure:1}a). The larger K\apex{+} ion (upper panel) 
leads instead to an incipient metal-to-insulator transition for 
large doping levels due to the introduction of disorder during the intercalation process. 
This disorder may originate from simple lattice distortions or a more complex 
coexistence of different incommensurate doped structures, such as those reported in superconducting 
intercalated TaS\ped{2} \cite{KashiharaJPSJ1981} and Bi\ped{2}Se\ped{3} \cite{HorPRL2010}.
These results demonstrate the critical importance of the specific ionic species and size in 
ion-gated devices, and indicate that different electrolytes can be used to 
explore different phase diagrams within the same material and device architecture.


We prepared few-layer MoS\ped{2} flakes by micromechanical exfoliation of 
their bulk crystals (2H polytype, SPI supplies) via
the well-known scotch-tape method \cite{FrindtPRL1972,BonaccorsoMatTod2012,Novoselov2005}
and transferred them on SiO\ped{2}($300$ nm)/Si substrates. We inspected 
the flakes with an optical microscope, and selected samples with the number
of layers between $\sim5$ and $10$ by analyzing their reflection 
contrast \cite{LiACSNano2016}. We realized the electrical contacts 
(Ti($5$ nm)/Au($35$ nm)) in Hall bar configuration, together with a co-planar 
side gate electrode, by standard microfabrication techniques. We patterned 
and deposited a solid oxide mask (Al\ped{2}O\ped{3} thickness $\sim 40$ nm) 
on the metallic leads only to reduce their 
interaction with the electrolyte during the experiments. 
Reactive Ion Etching (Ar gas, RF Power $100$ W, exposure time $2$ min) was used to pattern the flakes
into a rectangular shape, in order to achieve a well-defined aspect
ratio for sheet resistance measurements. 
Fig. \ref{figure:1}b presents the optical micrograph of a completed device
before drop-casting the polymer electrolyte prepared by dissolving $\sim 25$ wt\% 
of either K\apex{+} or Li\apex{+}-based salts in polyethylene glycol (PEG, $M_w \sim 600$). 
We tested both ClO\ped{4}\apex{-} and bis(trifluoromethane)sulfonimide(TFSI\apex{-})-based 
salts, and observed no significant dependence of the gating efficiency on the anion choice.
Both Li\apex{+} and K\apex{+} electrolytes were liquid at room
temperature and underwent a glass transition below $\sim 250$ K. Transport 
measurements were performed as a function of the temperature $T$ via the 
standard lock-in technique in a Quantum Design\textregistered{} Physical 
Properties Measurement System with minimal exposure to ambient condition.

We accessed the intercalated state in our MoS\ped{2} devices by slowly 
($dV_G/dt\sim2$ mV/s) ramping the gate voltage $V_G$ to a target value 
at $T = 300$ K and monitoring their conductivity for sharp increases 
in its value as the signature of the onset of intercalation \cite{YuNatureNano2015} 
(see Fig. \ref{figure:1}c). However, intercalation allows the ions 
in the electrolyte to migrate across the entire thickness of the device, and the increase 
in conductivity may potentially be suppressed by an increase in disorder. Hence, 
its onset can more reliably be detected as a large increase in
the Hall carrier density $n_{H} = 1/eR_{H}$ of the device to values comparable
with those of a few-nanometer-thick metal ($\sim10^{15}$ cm\apex{-2}). 
These values are one order of magnitude larger than those achievable 
on MoS\ped{2} upon pure surface accumulation \cite{YeScience2012,ShiSciRep2015,CostanzoNatureNano2015,BiscarasNatCommun2015} 
and are thus a reliable signature for the onset of bulk doping.

Thus, when the target $V_G$ was reached, we waited for $\sim 30$ minutes 
as sufficient time allowing the full relaxation of ion dynamics to 
improve doping homogeneity. We then cooled the sample to 
$T \lesssim 240$ K (below the glass transition of the electrolyte) and measured the 
Hall coefficient $R_{H}$ by sweeping the magnetic field perpendicular to the surface 
of the active channel. At this point, we either performed a full $T$-dependent characterization 
of the transport properties of the device by cooling the system down to $2$ K, or we warmed 
the sample up to $300$ K and increased $V_G$ even further. We performed the $T$-dependent
characterization both before (ionic-gating regime) and after (ionic-doping regime) 
the onset of intercalation on our devices.

Fig. \ref{figure:1}c shows a comparison between the $V_G$ dependence 
of the sheet conductivity $\sigma_s$ of four devices, two gated with 
the KClO\ped{4}/PEG electrolyte (devices A and B), the other two with the 
LiTFSI/PEG electrolyte (devices C and D). While the details of these bias 
ramps vary between different samples, the same choice of electrolyte results 
in similar curves across multiple devices. We attribute the random appearance of 
step features in $\sigma_s$ to the dynamics of the intercalation process: each step 
corresponds to a different doping state, and these states are sample-dependent. 
Moreover, the behavior of K\apex{+}- and Li\apex{+}-gated devices is clearly different. 

We first consider the behavior of a K\apex{+}-gated device (device A): 
in this case, the gate voltage was ramped up to a maximum of $+3.1$ V, 
and $R_H$ was measured twice: first at $V_G = +2.8$ V, 
and then at $V_G = +3.1$ V. The corresponding values of $n_{H}$ show 
that the carrier density at $V_G=+2.8$
V ($n_{H}\simeq4.4\times10^{14}$ cm\apex{-2}) is about
six times smaller than the one at $V_G = +3.1$ V
($n_{H}\simeq2.6\times10^{15}$ cm\apex{-2}). This strongly
suggests that the device is still mainly in the electrostatic
accumulation regime at $V_G = +2.8$ V, and is instead 
intercalated at $V_G = +3.1$ V. It is worth noting that
this large increase in $n_H$ does not lead to a significant
increase in $\sigma_s$, indicating that doping
with K\apex{+} ions, while inducing carriers, severely reduces 
the carrier mobility (at $T = 300$ K, $\mu_{H}\simeq12\pm3$ 
and $2.5\pm0.2$ cm\apex{2}/Vs for $V_G = +2.8$ and $+3.1$ V, 
respectively). We can also roughly estimate the nominal doping level
$x$ in the K$_x$MoS\ped{2} stoichiometry at $V_G = +3.1$ V 
(K\ped{0.45}MoS\ped{2}), assuming a uniform distribution of the dopants
in all the layers (five for this specific sample). This estimation indicates
that the sample at $V_G = +3.1$ V should be completely
in the metallic state, and in the correct doping range to show superconductivity
at low temperature \cite{WoollamMSE1977}. Inducing larger 
doping levels in K\apex{+}-gated devices by applying gate voltages 
in excess of $V_G = +3.5$ V always leads to device failure.

Let us focus now on the behavior of a Li\apex{+}-gated device (device C). 
Interestingly, Li\apex{+}-gated devices did not show significant signs of intercalation in
the same voltage range for which intercalation occurred in the K\apex{+}-gated devices. 
Instead, we observed an electrostatic increase of $\sigma_s$ with
increasing gate voltage up to $V_G\simeq+3.6$ V.
Larger voltage values caused a peculiar behavior to emerge, where
$\sigma_s$ appeared to randomly ``jump'' between high- and low-conductivity
states as $V_G$ was increased. This behavior, which may be associated 
with an unstable incorporation of the Li\apex{+} ions between the MoS\ped{2} 
layers, continued up to $V_G\simeq+6.1$ V. Even larger gate voltages
up to $V_G\simeq+7.0$ V featured a second stable
region of monotonically increasing $\sigma_s$, which was about $4$
times larger than that for $V_G\simeq+3.6$ V.
The corresponding values of carrier density, as measured by Hall effect 
at $T = 220$ K, were $n_{H}(+3.6\,V)\simeq2.8\times10^{14}$ cm\apex{-2} 
and $n_{H}(+7.0\,V)\simeq7.1\times10^{14}$ cm\apex{-2}
(Li\ped{0.12}MoS\ped{2}), with a Hall mobility 
$\mu_{H}\simeq12\pm2$ and $9.2\pm1.8$ cm\apex{2}/Vs
in the two cases, respectively. The significant increase in both 
$\sigma_s$ and $n_H$ indicate that the high-conductivity 
state at $V_G\simeq+7.0$ V may be associated with Li\apex{+} intercalation. 
The corresponding nominal doping $x\simeq 0.12$ achieved in our sample 
is still below the onset of superconductivity in chemically intercalated samples, 
which emerges only for $x\geq 0.4$ \cite{WoollamMSE1977}. 

Overall, the following main differences emerge when comparing K\apex{+}
and Li\apex{+} intercalation at the same operating temperature 
($T = 300$ K): first, the decrease in mobility is much less 
pronounced in the case of Li\apex{+} doping, indicating 
a much less prevalent introduction of extra defects in the system; 
second, while the thickness of the two samples was 
comparable, the final $n_{H}$ is significantly smaller in the
Li\apex{+}-doped one, indicating that K\apex{+} ions are able to more 
efficiently penetrate between the MoS\ped{2} layers. Furthermore, 
the onset of K\apex{+} doping requires smaller gate voltages, 
but leads to device degradation for smaller $V_G$ values as well.

\begin{figure}
\includegraphics[width=0.8\columnwidth]{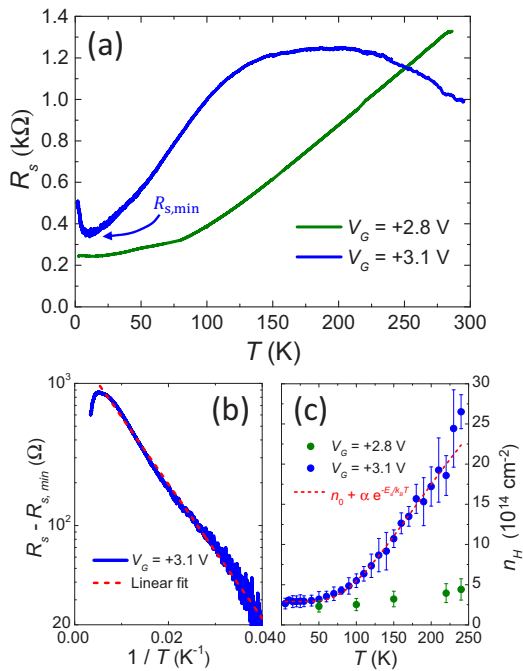}
\caption{$T$-dependent transport properties of K\apex{+}-gated MoS\ped{2}. (a) $R_s\,\mathrm{vs.}\,T$ for K\apex{+} accumulation (solid green line) and intercalation (solid blue line). (b) $R_s - R_{s,min}$ as a function of $T^{-1}$ in the intercalated state. Dashed red line is a linear fit to the curve to highlight its exponential dependence. (c) $n_H\mathrm{\,vs.}\,T$ corresponding to the curves of (a). Dashed red line is a fit to the thermally-activated behavior.}
\label{figure:2}
\end{figure}

We now consider the $T$-dependent transport properties of our devices down to $2$ K 
in both K\apex{+} and Li\apex{+}-doped samples. We characterize our devices first 
in the electrostatic regime, and again after the electric field has driven the ions 
to intercalate the material. 

Fig. \ref{figure:2}a shows the $T$-dependence of the square resistance $R_s$ of 
device A, gated with the KClO\ped{4}/PEG electrolyte, for both ionic gating 
($V_G = +2.8$ V, green curve) and ionic doping 
($V_G = +3.1$ V, blue curve). When the ions only 
accumulate at the surface of MoS\ped{2} (low $V_G$), the device shows a clear metallic 
behavior, with a smaller low-$T$ value of $R_s$ than that 
typically displayed by ionic-liquid-gated MoS\ped{2} \cite{YeScience2012}. 
This is consistent with the larger doping level induced in the sample. Moreover, 
this suggests that K\apex{+} gating is able to bring MoS\ped{2} beyond 
the field-induced superconducting dome \cite{YeScience2012}.

When the ions are able to intercalate the sample, we would also expect a metallic 
behavior and a further reduction of $R_s$ at low-$T$. Moreover, 
given that the doping level K\ped{0.45}MoS\ped{2} determined at $240$ K, we 
would also expect the emergence of a superconducting transition 
at $T\sim6$ K \cite{SomoanoJCP1973}. However, the $T$-dependence of 
$R_s$ in the intercalated state does not show 
any of these features. Instead, it shows a clear non-monotonic behavior 
and two regions where $R_s$ decreases for increasing $T$: one for
$T\gtrsim150$ K and one for $T\lesssim20$ K. The second one,
the low-temperature upturn, is insensitive to the applied magnetic
field, ruling out a possible contribution from weak localization.
For intermediate temperatures, $R_s$ increases 
as $e^{-A/T}$, $A\simeq 107$ K (see Fig. \ref{figure:2}b). 
This type of behavior is reminiscent of a two-dimensional system 
very close to a metal-to-insulator transition \cite{HaneinPRL1998,MeirPRL1999}.

These results indicate the peculiar condition of a system being
close to becoming an insulator, while at the same time presenting
a metal-like density of charge carriers at high $T$. Thus, we investigated 
whether $n_H$ was metallic at low-$T$ as well. 
Fig. \ref{figure:2}c shows the $T$-dependence of $n_{H}$ obtained from 
Hall effect measurements. It is apparent that $n_{H}$ in the bulk doped 
state (blue dots) strongly decreases at the reduction of $T$. 
Indeed, the $T$-dependence of $n_{H}$ can be separated 
into two contributions: a relatively small constant
value $n_{0}\simeq2.9\times10^{14}$ cm\apex{-2} and an
Arrhenius-like term $n(T)\propto e^{-E_{a}/k_{B}T}$, where $E_{a}\simeq0.03$ eV 
is an activation energy and $k_{B}$ is the Boltzmann constant. 
For comparison, the carrier density induced by surface ionic gating 
(green dots) is much less $T$-dependent, while at the same time reaching 
nearly the same low-$T$ value. The resulting low-$T$ mobilities are $\mu_{H}\simeq110\pm33$ 
and $50\pm12$ cm\apex{2}/Vs for K\apex{+} accumulation 
and intercalation respectively. Thus, it is natural to assume that the 
quasi-constant term arises from ionic gating at the sample surface, while the 
thermally-activated one is associated with bulk ion doping. 

We thus suggest that the electrochemically intercalated K\apex{+} ions are
behaving as thermally-activated electron donors and reside
in shallow trap states in the bulk MoS\ped{2} energy gap:
the material thus behaves more like a highly-doped but highly-defective
semiconductor with a field-induced metallic channel at its surface,
instead of showing a proper metallic character across its entire thickness.
Moreover, this very defective character of the K\apex{+}-doped regime 
is able to account for both the sharp reduction in carrier mobility, and 
the emergence of an Anderson-like localization regime at low $T$. 
A disorder-induced metal-to-insulator transition was recently reported 
in ion-gated monolayer ReS\ped{2} \cite{OvchinnikovNComms2016}, 
but not in any ion-gated multilayer transition metal dichalcogenide.

\begin{figure}
\includegraphics[width=0.8\columnwidth]{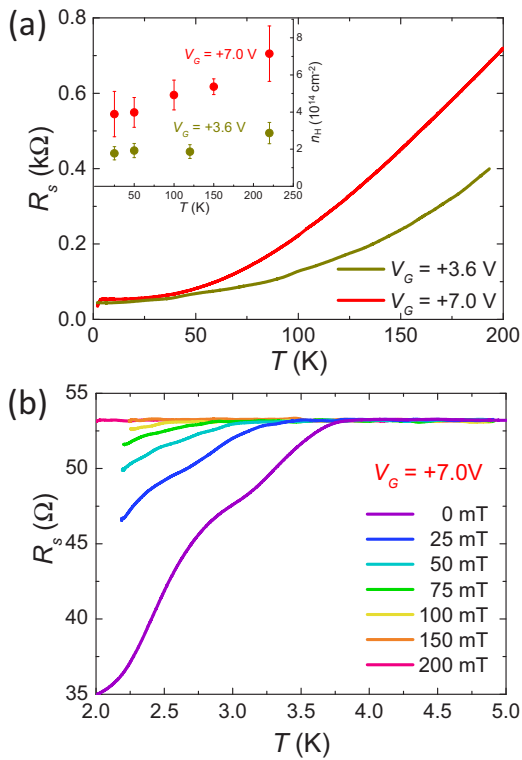}
\caption{$T$-dependent transport properties of Li\apex{+}-gated MoS$\ped{2}$. (a) $R_s\,\mathrm{vs.}\,T$ for Li\apex{+} accumulation (solid yellow line) and intercalation (solid red line). Inset shows the corresponding carrier densities $n_H$. (b) $R_s\,\mathrm{vs.}\,T$ in the intercalated state below $5$ K for different values of the applied magnetic field.}
\label{figure:3}
\end{figure}

In Fig. \ref{figure:3}a instead we present the $R_s\,\mathrm{vs.}\,T$ behavior of 
device C, gated with the LiTFSI/PEG electrolyte. The yellow 
and red curves refer to Li\apex{+}-gating ($V_G = +3.6$ V) 
and doping ($V_G = +7.0$ V) respectively. 
The inset shows the corresponding $T$-dependence of their 
sheet carrier density $n_{H}$ as measured by Hall effect. Unlike 
the K\apex{+} ion, the Li\apex{+} ion allows the system to retain 
a full metallic behavior also in the bulk doping regime, 
without evidences of non-monotonicity or low-$T$ upturns. The $T$-dependence
of $n_{H}$ is also less pronounced, being nearly constant for
$T\lesssim150$ K in the case of ionic gating and losing
less than half of its high-$T$ value in the case of ionic doping. 
Indeed, the low-$T$ carrier density in the Li\apex{+}-doped 
state, $n_{H}\simeq3.9\times10^{14}$ cm\apex{-2}, was significantly
larger than the one for K\apex{+} doping, even though its nominal 
doping level $x$ was nearly $3$ times smaller. This indicates that, in the case 
of Li\apex{+} doping, the smaller density of defects acting as shallow 
trap states allows for a higher fraction of charge carriers to
participate in conduction at low $T$. This reduced density of defects 
is also apparent in the low-$T$ mobilities $\mu_{H}\simeq800\pm160$ 
and $300\pm94$ cm\apex{2}/Vs for Li\apex{+} gating  
and doping respectively, several times larger than the ones 
we observed in the case of the K\apex{+} ion. 

%
The most likely explanation of these results is that the size of
the K\apex{+} ion is too large to be able to intercalate
the MoS\ped{2} lattice without introducing significant
distortions and defects in its entire volume. These defects would
then act as shallow trap states, capturing most of the transferred
electrons at low $T$ and suppressing the metallic
behavior except in the thin layer at the surface due to electrostatic
accumulation. We note that a similar disruptive effect of large intercalating species 
was also observed in ion-gated TaS$\ped{2}$, where it leads to abrupt 
device failure \cite{YuNatureNano2015}. It is interesting then to consider 
why the larger K\apex{+} ion shows an enhanced doping efficiency with respect to 
the smaller Li\apex{+}. We suggest that this behavior may arise from the lattice distortions 
introduced during the intercalation process allowing the K\apex{+} ions still 
dissolved in the electrolyte to diffuse more easily through the damaged regions. 
On the other hand, the lattice remains relatively unaffected during the intercalation 
by the smaller Li\apex{+} ions, thus requiring larger driving voltages to 
intercalate the bulk of the sample. However, further investigations 
- such as disorder studies by means of x-ray diffraction - 
are needed to clarify this issue.

Further evidence of the importance of dopant size on the behavior 
of ion-gated devices lies in the fact that we were able to observe 
a clear downturn in the $R_s\,\mathrm{vs.}\,T$ 
curve in the Li\apex{+}-doped state below $4$ K. Fig. \ref{figure:3}b 
shows its response to the application of a magnetic field
perpendicular to the active channel of the device. While the downturn 
never reaches a zero-resistance state, its suppression by a magnetic field 
is precisely the behavior expected from a superconducting
transition. We point out that while the nominal doping level 
at $V_G=+7.0$ V was estimated to be Li\ped{0.12}MoS\ped{2}, 
the onset temperature of the downturn ($T_{c}^{on}\simeq3.7$
K) agrees well with that of chemically doped Li$_x$MoS\ped{2} for $x\geq0.4$
\cite{SomoanoJCP1973}. Moreover, superconductivity does not appear in chemically doped 
Li$_x$MoS\ped{2} for $x\leq0.4$ \cite{SomoanoJCP1973}. 
Since we observe a superconducting onset, the doping level in the 
intercalated state must be strongly inhomogeneous. This is supported 
by the behavior of the superconducting transition: the $R_s\,\mathrm{vs.}\,T$ 
profile is not the sharp drop associated with homogeneous bulk superconductivity. Instead,
the transition is broad and strongly suggestive 
of multiple phases. This kind of behavior is typical of
granular superconductors: in the Li\apex{+}-doped state only a handful of
regions are able to reach a doping level large enough to induce a superconducting
state, while most of the active channel remains metallic and prevents the 
realization of homogeneous $3$D superconductivity. The slowly vanishing resistance tail 
is due to Josephson tunneling between the superconducting regions 
(weak-link superconductivity) \cite{LikharevRMP1979,ClaassenAPL1980}. 


In conclusion, we employed polymer electrolyte gating to intercalate 
MoS\ped{2} thin flakes with different ionic species. We unveiled 
the critical role of ionic size in the determination of the electric 
transport properties of the intercalated devices. The larger K\apex{+} 
ions were found to strongly damage the MoS\ped{2} lattice leading to 
an incipient metal-to-insulator transition at high doping levels. The 
smaller Li\apex{+} ions preserved the metallic character of the 
devices and allowed the emergence of an inhomogeneous bulk 
superconducting phase. These findings highlight the critical role 
of the ionic medium in electrochemically gated devices, both 
for electrostatic carrier accumulation and field-driven ion intercalation.

\section*{Supplementary Material}
See Supplementary Material for further details on the measurement setup, Hall effect measurements, and optical characterization of the intercalation process.

\begin{acknowledgments}
We thank R. S. Gonnelli for perusing the manuscript and useful scientific discussions. 
We acknowledge funding from the European Research Council (Consolidator Grant no. 648855 Ig-QPD).
\end{acknowledgments}

\end{document}